\documentclass[
10pt,
twocolumn,
 amsmath,amssymb,
 aps,
 superscriptaddress,
]{revtex4-2}

\usepackage{graphicx}
\usepackage{dcolumn}
\usepackage{bm}
\usepackage{physics}
\usepackage[T1]{fontenc}

\begin{document}

\preprint{APS/123-QED}

\title{Effects of electronic correlation on the high harmonic generation in helium: \\ a time-dependent configuration interaction singles vs time-dependent full configuration interaction study}

\author{Aleksander P. Wo\'zniak}
\email{awozniak@chem.uw.edu.pl}
\affiliation{Faculty of Chemistry, University of Warsaw, Pasteura 1, 02-093 Warsaw, Poland}

\author{Micha{\l} Przybytek}
\affiliation{Faculty of Chemistry, University of Warsaw, Pasteura 1, 02-093 Warsaw, Poland}

\author{Maciej Lewenstein}
\affiliation{ICFO - Institut de Ciencies Fotoniques, The Barcelona Institute of Science and Technology, Av. Carl Friedrich Gauss 3, 08860 Castelldefels (Barcelona), Spain}
\affiliation{ICREA, Pg. Llu\'is Companys 23, 08010 Barcelona, Spain}

\author{Robert Moszy\'nski}
\affiliation{Faculty of Chemistry, University of Warsaw, Pasteura 1, 02-093 Warsaw, Poland}

\date{\today}

\begin{abstract}
In this paper, we investigate the effects of full electronic correlation on the high harmonic generation in the helium atom subjected to laser pulses of extremely high intensity. To do this, we perform real-time propagations of the helium atom wavefunction using quantum chemistry methods coupled to Gaussian basis sets. The calculations are done within the real-time time-dependent configuration interaction framework, at two levels of theory: time-dependent configuration interation with single excitations (TD-CIS, uncorrelated method) and time-dependent full configuration interaction (TD-FCI, fully correlated method), and analyse obtained HHG spectra. The electronic wavefunction is expanded in Dunning basis sets supplemented with functions adapted to describing highly excited continuum states. We also compare the TD-CI results with grid-based propagations of the helium atom within the single-active-electron approximation. Our results show when including the dynamical electron correlation, a noticeable improvement to the description of HHG can be achieved, in terms of e.g. a more constant intensity in the lower energy part of the harmonic plateau. However, such effects can be captured only if the basis set used suffices to reproduce the most basic features, such as the HHG cutoff position, at the uncorrelated level of theory.
\end{abstract}

\maketitle


\section{\label{sec:intro}Introduction}

Attoscience is a rapidly developing area of physics that studies light-matter interactions of atoms and molecules subjected to intense laser fields. One of the key aspects of attoscience is the high harmonic generation (HHG), which provides ultrashort pulses of coherent radiation of wavelength ranging from visible light to soft X-ray \cite{agostini2004,corkum2007,midorikawa2011}. Such pulses have recently became an extremely powerful tool for studying electronic dynamics in atoms and molecules\cite{bruner2016,nisoli2017,ossiander2017,marciniak2019}, charge transfer reactions \cite{tuthill2020}, molecular imaging \cite{torres2010,vozzi2011,peng2019} and many more. According to the famous three-step model \cite{corkum1993,kulander1993,lewenstein1994}, during the HHG process the electron is (i) removed from the atom or molecule by the strong laser field via tunneling ionization, (ii) accelerated away in the field until the field sign is changed and then reaccelerated towards the residual ion, and (iii) recombined with the ion, which is accompanied by the production of a high energy photon. The HHG spectrum is characterized by a long plateau region, composed of consecutive harmonic peaks, which ends with an abrupt cutoff at the energy $E_\textrm{cut}$ connected to maximum kinetic energy an electron can gain in the electromagnetic field.

HHG has been theoretically studied for decades using a variety of methods \cite{eden2004,spanner2013,amini2019,yu2019,coccia2022}. Over the last few years, as both the theoretical and experimental attoscience shifts towards increasingly more complex systems including solids \cite{ghimire2011,hansen2018} and biomolecules \cite{marangos2016,harada2018,baykusheva2019}, the real-time propagation of the wavefunction using time-dependent quantum chemical methods coupled to $L^2$-integrable basis sets have gained some recognition in describing the electron dynamics during attosecond processes \cite{ishikawa2015,goings2017,saalfrank2020,bedurke2021,coccia2022}. Due to a better scaling with the number of electrons and much easier handling of multicenter systems, they are slowly surpassing the purely numerical, grid-based approaches.

The approaches known from quantum chemistry that have been extended to the time-domain can be generally classified into two categories: the orbital-based ones, which include time-dependent Hartree-Fock (TD-HF) \cite{kulander1987,kulander1988,pindzola1991,pindzola1995,li2005,nikolopoulos2007,ding2011} and time-dependent density functional theory (TD-DFT) \cite{runge1984,tong1998,castro2004,chu2011,ding2011,lehtovaara2011,lopata2011,redkin2011,sonk2011b,luppi2012,castro2015,kuisma2015,wopperer2015,provorse2016,sissay2016,gao2017,rossi2017,vincendon2017,wu2017,parise2018,pemmaraju2018,wu2018,reiff2020,tancogne2020,bedurke2021,pauletti2021}, and the wavefunction-based ones, such as time-dependent configuration interaction (TD-CI) \cite{klamroth2003,huber2005,krause2005,rohringer2006,krause2007,schlegel2007,tremblay2008,klinkusch2009a,klinkusch2009b,greenman2010,sonk2011a,hochstuhl2012,luppi2012,ramakrishnan2013,bauch2014,hochstuhl2014,krause2014a,krause2014b,alcoba2016,pabst2016,white2016,artemyev2017a,artemyev2017b,sato2018a,bedurke2019,coccia2019,coccia2020,lee2020,saalfrank2020,bedurke2021,luppi2021,pauletti2021}, time-dependent multiconfigurational self consistent field (TD-MCSCF) \cite{kitzler2004,caillat2005,nest2007,miyagi2013,sato2013,hochstuhl2014,sato2015,sato2016,sawada2016} and time-dependent coupled cluster (TD-CC) \cite{huber2011,kvaal2012,sato2018b,pedersen2019,kristiansen2020,pathak2020a,pathak2020b,pathak2021}. Out of all these methods the one that gained an exceptional popularity in theoretical attoscience is the time-dependent configuration interaction with single excitations (TD-CIS) \cite{klamroth2003,huber2005,rohringer2006,krause2007,schlegel2007,klinkusch2009a,klinkusch2009b,greenman2010,sonk2011a,luppi2012,krause2014a,krause2014b,pabst2016,white2016,sato2018a,bedurke2019,coccia2019,coccia2020,lee2020,saalfrank2020,bedurke2021,luppi2021,pauletti2021}, due to its simple formalism and low computational costs. In TD-CIS the time-dependent wavefunction is expanded in (time-independent) CIS eigenstates obtained via diagonalization of an atomic or molecular Hamiltonian matrix that consists of matrix elements between the Hartree-Fock ground state configuration and singly excited configurations (the TD-CI formalism will be detailed more comprehensively in the next section). TD-CIS has proven useful in predicting the HHG spectra in molecules as large as chlorinated hydrocarbons \cite{bedurke2019} and nucleobases \cite{luppi2021}. However, since the CIS method is an equivalent of the Hartree-Fock method for excited states, it completely lacks the electronic correlation effects, that are considered by many to play an important role in laser-driven electron dynamics \cite{shiner2011,pabst2013,neufeld2020}. A natural way to account for these effects would be to include higher excitations in the CI wavefunction, starting with CI with single and double excitations (CISD). The TD-CISD approach has been discussed several times in the literature \cite{krause2005,coccia2017} and tested for two-electron systems in small basis sets \cite{krause2007,schlegel2007,saalfrank2020}, but is hardly applied in the strong field calculations due for reasons. First, inclusion of doubly excited configurations in the CI Hamiltonian matrix drastically increases its size, making both the obtainment of the CI eigenstates and the real-time propagation far more expensive and time consuming, if feasible at all. Secondly, many authors base their real-time propagation codes on results from existing electronic structure packages, such as Gaussian \cite{krause2014a,krause2014b}, GAMESS \cite{coccia2020} or Q-Chem \cite{luppi2012,white2016,coccia2019}. Most of these programs usually perform the diagonalization of large matrices using iterative algorithms, that provide no more than the ground and a few excited electronic states, while the real-time propagation usually requires a full Hamiltonian eigenspectrum.

Nevertheless, there have been some other attempts to account for the electronic correlation effects in HHG. A notable example is the development of the TD-CIS(D) method, which treats the effects of double excitations via perturbational corrections to CIS energies \cite{krause2005,krause2007,schlegel2007,tremblay2008,sonk2011a,luppi2012}. Unfortunately, as reported by Sonk \textit{et al.} \cite{sonk2011a}, it may give an erratic optical response in intense fields. Probably the most common way relies on using TD-DFT methods with an exchange-correlation potential of choice \cite{chu2011,ding2011,lopata2011,redkin2011,sonk2011b,luppi2012,castro2015,kuisma2015,sissay2016,gao2017,reiff2020,bedurke2021,pauletti2021}. However, as it was discussed, first by Pindzola \textit{et al.} \cite{pindzola1991} and later eg. by Sato and Takeshi \cite{sato2014}, a single determinant wavefuction, whether constructed from Hartree-Fock or Kohn-Sham spinorbitals, is inherently unable to describe the single ionization processes, especially in closed-shell systems. This is due to the fact that if two spinorbitals in the ground state share the same spatial component and there is no spin-dependent perturbation in the time-dependent operator representing the interaction between electrons and the external laser field (as it is usually assumed), they will always evolve in the same manner in order to maintain the closed-shell structure, making it impossible to represent a state in which one electron becomes ionized and the other remains in the ionic core. Since tunneling ionization is an intrinsic part of HHG and may compete with above threshold and barrier suppression ionization under certain laser conditions, this may give an incorrect physical picture of the whole process. The way to overcome this issue is either to  use a spin-dependent external field that enforces different spatial movement of electrons of different spin \cite{isborn2009}, or turn to multideterminant correlated methods like time-dependent coupled cluster, time-dependent algebraic diagrammatic construction (TD-ADC) \cite{ruberti2014,ruberti2018} and the above-mentioned TD-CISD.

In this work, we explore the applicability of the TD-CISD approach for the calculation of the HHG spectra, aiming at answering the question whether the explicit inclusion of the electronic correlation effects brings a substantial improvement to the description of the harmonic generation process, that can outweigh the cumbersome computational costs of this method.
In order to do this, we perform the real-time TD-CISD propagations of the helium atom wavefunction, which serves us as a model multi-electron system. From the propagations we compute the HHG spectra and compare them to analogous results obtained using the TD-CIS method and the grid-based single-active-electron calculations.
The CI helium wavefunctions are expanded in single-electron atom-centered Gaussian basis sets of different sizes.
This also allows us to a address another question of which way of enhancing the time-dependent wavefunction expansion -- by augmenting the configurational basis through the addition of double excitations or by increasing the orbital basis by adding more Gaussian functons -- provides a better improvement to the description of HHG.
Choosing a two-electron atom has also an additional advantage, as for such a system CISD is equivalent to full CI (no excitations higher than double can be generated). This gives us the opportunity to compare results obtained using a completely uncorrelated method with the ones from a method that provides an exact correlation energy within a given basis set.

This paper is organised as follows.
In section 2 we provide a brief theoretical background of the applied methods and discuss the computational details.
Then in section 3 we present and analyze our results.
Finally, in section 4 we summarize our work and provide some outlook for future studies.

\section{Methods}

\subsection{The time-dependent configuration interaction formalism}

We solve the time-dependent Schr\"odinger equation (TDSE) for the electronic wavefunction of the helium atom subjected to intense laser field (the Hartree atomic units are used throughout the paper unless stated otherwise)
\begin{equation}
i \frac{\partial}{\partial t}\vert\Psi(\mathbf{r},t)\rangle = (\hat H_0+\hat V(t))\vert\Psi(\mathbf{r},t)\rangle,
\end{equation}
where $\hat H_0$ is the time independent electronic Hamiltonian in the Born-Oppenheimer approximation (assuming the position of the nucleus at the center of the coordinate system),
\begin{equation}
\hat H_0(\vec r_1,\vec r_2)=-\frac{\nabla_1^2}{2}-\frac{\nabla_2^2}{2}-\frac{2}{r_1}-\frac{2}{r_2}-\frac{1}{\lvert \vec r_1 - \vec r_2 \rvert},
\end{equation}
and $\hat V(t)$ is the time-dependent operator coupling the electrons to the external field.
Within the real-time TD-CI approach the time-dependent wavefunction  is written as a linear combination of the time-independent CI states (appoximate eigenstates of $\hat H_0$)
\begin{equation}
\ket{\Psi(t)} = \sum_k c_k(t) \ket{\Psi^\textrm{CI}_k} 
\end{equation}
Each CI state can be expanded in the basis constisting of the ground state (reference) Slater determinant $\Phi_0$ built from occupied spinorbitals obtained through the solution of Hartree-Fock equations within a chosen basis set, and the set of excited determinants, $\{\Phi_{\nu}^{\gamma},\Phi_{\nu\mu}^{\gamma\delta},...\}$ that are created from the reference determinant by replacing a certain number of occupied spinorbitals $\{\nu,\mu,...\}$ with virtual spinorbitals $\{\gamma,\delta,...\}$.
The maximum number of replaced spinorbitals is one for CIS states, two for CISD states, three for CISDT states etc.
However, in case of the real-time TD-CI calculations the wavefunction is bound to evolve only on the manifold of states with spin equal to the ground state spin, so it is more convenient to express the CI states in the basis of configuration state functions (CSFs) -- linear combinations of Slater determinants, which are eigenfunctions of the square of the total spin operator.
This allows for a significant reduction of the necessary CI space.
For a closed-shell system, such as the helium atom, besides the HF reference determinant, we have one type of CIS singlet excited CSFs (the roman letters denote the spatial orbitals, and the bars indicate whether we replace an $\alpha$ or a $\beta$ spinorbital associated with a given orbital) \cite{szabo2012,krause2007},
\begin{eqnarray}
&&^\mathrm{1}\Phi_i^a = \frac{1}{\sqrt{2}} (\Psi_i^a + \Psi_{\bar{i}}^{\bar{a}}),
\end{eqnarray}
and five types of CISD singlet excited CSFs,
\begin{eqnarray}
&&^\mathrm{1}\Phi_{ii}^{aa} = \Phi_{i \bar{i}}^{a \bar{a}}, \nonumber \\
&&^\mathrm{1}\Phi_{ii}^{ab} = \frac{1}{\sqrt{2}} (\Phi_{i \bar{i}}^{a \bar{b}} + \Phi_{\bar{i} i}^{\bar{a} b}), \nonumber \\
&&^\mathrm{1}\Phi_{ij}^{aa} = \frac{1}{\sqrt{2}} (\Phi_{i \bar{j}}^{a \bar{a}} + \Phi_{\bar{i} j}^{\bar{a} a}), \\
&&^\mathrm{A}\Phi_{ij}^{ab} = \frac{1}{\sqrt{12}} (2\Phi_{ij}^{ab} + 2\Phi_{\bar{i}\bar{j}}^{\bar{a}\bar{b}} + \Phi_{i \bar{j}}^{a \bar{b}} + \Phi_{\bar{i} j}^{\bar{a} b} - \Phi_{i \bar{j}}^{b \bar{a}} - \Phi_{\bar{i} j}^{\bar{b} a}), \nonumber \\
&&^\mathrm{B}\Phi_{ij}^{ab} = \frac{1}{2} (\Phi_{i \bar{j}}^{a \bar{b}} + \Phi_{\bar{i} j}^{\bar{a} b} + \Phi_{i \bar{j}}^{b \bar{a}} + \Phi_{\bar{i} j}^{\bar{b} a}),\nonumber
\end{eqnarray}
where $i>j$, $a>b$, so the total expressions for CIS and CISD states read
\begin{eqnarray}
\Psi^\textrm{CIS}_k = &&C_{0,k} \Phi_0 + \sum_{i,a} C^a_{i,k} {^\mathrm{1}}\Phi_i^a, \\
\Psi^\textrm{CISD}_k = &&C_{0,k} \Phi_0 + \sum_{i,a} (C^a_{i,k} {^\mathrm{1}}\Phi_i^a + C^{aa}_{ii,k} {^\mathrm{1}}\Phi_{ii}^{aa}) \nonumber \\
&& + \sum_{i,a,b} C^{ab}_{ii,k} {^\mathrm{1}}\Phi_{ii}^{ab} + \sum_{i,j,a} C^{aa}_{ij,k} {^\mathrm{1}}\Phi_{ij}^{aa} \nonumber \\
&& + \sum_{i,j,a,b} ({^\mathrm{A}}C^{ab}_{ij,k} {^\mathrm{A}}\Phi_{ii}^{ab} + {^\mathrm{B}}C^{ab}_{ij,k} {^\mathrm{B}}\Phi_{ii}^{ab}).
\end{eqnarray}
These are general expressions valid for an arbitrary closed-shell system, however is is worth noting that in case of the helium atom the CSF types $\Phi_{ij}^{aa}$, $^A\Phi_{ij}^{ab}$ and $^B\Phi_{ij}^{ab}$ are absent from the second expansion, since there is only one occupied spatial orbital in the HF ground state.
The coefficients $C_k$ and the eigenenergies of corresponding CI states $E_k^{\textrm{CI}}$ are determined by diagonalizing the matrix of the $H_0$ operator in the basis of all CSFs available at the given level of theory,
\begin{equation}
\mathbf{H_0^\textrm{CI}}\mathbf{C}_{k} = E_k^{\textrm{CI}}\mathbf{C}_{k}
\end{equation}
The expressions for the matrix elements of $\mathbf{H_0^\textrm{CIS}}$ and $\mathbf{H_0^\textrm{CISD}}$ are provided in the supplementary material.

\subsection{Real-time propagations}
In our calculations the laser pulse is represented by a linearly polarized oscillating electric field with a sine-squared envelope of shape
\begin{equation} \label{eq:efield}
E(t) = 
\begin{cases}
E_0 \sin(\omega_0 t) \sin^2(\omega_0 t/2 n_c)&\text{if}\; 0 \leq t \leq 2\pi n_c/\omega_0, \\
0& \text{otherwise,}
\end{cases}
\end{equation}
with the carrier frequency $\omega_0$ = 1.55 eV ($\lambda_0$ = 800 nm), the number of optical cycles $n_c$ = 20, and the total pulse duration $2\pi n_c/\omega_0 \approx$ 2206 a.u. ($\approx$ 53.4 fs). The maximum field amplitude $E_0$ is related to the laser intensity $I_0$ via $I_0=\varepsilon_0cE_0^2/2$. We examine three laser intensities: $2 \times 10^{14}$ W/cm\textsuperscript{2}, $3 \times 10^{14}$ W/cm\textsuperscript{2} and $5 \times 10^{14}$ W/cm\textsuperscript{2}. Their choice was based mainly on the value of the Keldysh parameter $\gamma = \sqrt{I_p/2U_p}$ \cite{keldysh1965}, where $I_p$ is the first ionization energy of helium and $U_p = E_0^2/4\omega_0^2$ is the ponderomotive energy of a free electron in the electromagnetic field. For the investigated laser intensities the Keldysh parameter equals 1.01, 0.83 and 0.64, respectively, indicating the tunneling regime of ionization, which is desirable for an effective harmonic generation process.

The external electric field is treated in the dipole approximation and the field coupling operator $V(t)$ is expressed in the length gauge (assuming that the field is polarized along the $z$-axis),
\begin{equation}
\hat V(t) = -\hat z E(t).
\end{equation}
The wavefunction is propagated using the middle-step Crank-Nicolson algorithm,
\begin{eqnarray}
    \left(\mathbf{I}+\frac{i\Delta t}{2}\mathbf{H^\textrm{CI}}(t+\Delta t/2)\right)\mathbf{c}(t+\Delta t) = \nonumber\\
    = \left(\mathbf{I}-\frac{i\Delta t}{2} \mathbf{H^\textrm{CI}}(t+\Delta t/2)\right)\mathbf{c}(t),
\end{eqnarray}
where $\mathbf{H^\textrm{CI}}(t)$ is the time-dependent Hamiltonian matrix (and $\mathbf{\textrm{CI}}$ refers to either $\mathbf{\textrm{CIS}}$ or $\mathbf{\textrm{CISD}}$), $\mathbf{I}$ is the identity matrix and $\mathbf{c}(t)$ is the vector of the time-dependent coefficients $c_k(t)$. Since the time-dependent wavefunction is expanded in the basis of CI eigenstates, the time-dependent Hamiltonian matrix simplifies to $\mathbf{H^\textrm{CI}}(t)$ = $\mathbf{E^\textrm{CI}} + \mathbf{V}(t)$, where $\mathbf{E^\textrm{CI}}$ is the diagonal matrix of CI eigenenergies. We start the propagation from the ground state of helium (calculated at either CIS or CISD level of theory, depending on the propagation type) and propagate the system in time for twice the duration of the laser pulse, with timestep $\Delta t$ = 0.01 a.u.

\subsection{Gaussian basis sets}

Difficulties in the proper representation of Rydberg and continuum electronic states are the major drawback of modeling the laser-driven electron dynamics using the basis set approach. These states are usually highly diffuse, so approximating them with a set of nuclei-centered functions may be troublesome. Some recent works, however, have shown \cite{luppi2013,coccia2016a,coccia2016b,labeye2018} that a good semi-quantitative description of HHG can be achieved when using a special type of Gaussian basis sets consisting of two types of functions. The first type is some standard basis set routinely used in quantum chemical calculations of ground-state properties, such as one of the Dunning basis sets \cite{dunning1989}. It comprises functions that can well reproduce the electronic ground state and some of the lowest-lying excited states, but fails to depict the motion of electrons far from the nuclei. Therefore it must be supplemented with the second type of functions designed to describe the highly excited and continuum states. Several types of Gaussian functions have been proposed for this second type, including functions developed by Kaufmann \textit{et al.} \cite{kaufmann1989,coccia2016a,coccia2016b,labeye2018} and even-tempered Gaussians \cite{white2016,bedurke2019}.

In our recent paper \cite{wozniak2021} we introduced a novel class of Gaussian functions we dubbed ARO functions, that proved able to approximate a wide spectrum of excited states necessary for the description of HHG in both one- and two-electron systems. Therefore, in the present calculations we use Dunning basis sets mixed with ARO functions. In order to select an optimal basis set for the time-dependent calculations we have performed a series of test propagations of the helium atom using different augmented correlation-consistent basis sets of general type x-aug-cc-pVYZ, with x=$\emptyset$,d,3,4 and Y=D,T,Q,5, supplemented with shells of ARO functions with angular momenta $l$ from 0 ($s$-type functions) up to $l_\textrm{max}$ = 2, 3 or 4 ($d$-,$f$- and $g$-type functions, respectively). The ARO functions were generated according to the procedure described in \cite{wozniak2021}, by fitting Gaussians to a set of Slater-type orbitals with exponent $\zeta$ = 1.7 (the effective nuclear charge of helium) and principal quantum numbers $n$ from 2 to 90. The number of ARO fuctions added per angular momentum varied from 4 to 10. The test calculations were performed at the TD-CIS level of theory, for all examined laser conditions. From the results of the propagations the HHG spectra were computed using the methodology described below and assessed by comparing their key features, including the distinguishability of harmonic peaks, the cutoff position and presence of numerical artifacts. The properties of the basis sets themselves, such as lack of linear dependencies between the functions, were also taken into account. Overall, the most reliable results were obtained by using basis sets based on the d-aug-cc-pVQZ Dunning basis set, out of which two were selected for the proper TD-CI calculations. The first one includes 8 ARO functions per each angular momentum from 0 to 4 and is dubbed d-aug-cc-pVQZ+ARO8g. It provided the most optimal spectra at the TD-CIS level, however its size -- 262 functions in total -- makes it too large to use in the TD-CISD calculations (the dimension of the CIS matrix expressed in singlet CSFs is $ov +1$, where $o$ and $v$ are the numbers of occupied and virtual orbitals, respectively, while the dimension of the CISD matrix is $\frac{1}{2}(o^2v^2+3ov+2)$). This is why we also use its smaller variant containing 6 ARO functions per each angular momentum from 0 to 3 (158 functions in total, named d-aug-cc-pVQZ+ARO6f), in which we perform both the TD-CIS and TD-CISD calculations.

We are aware of the fact that ideally the basis set should be optimized (in terms of number and range of the ARO functions) separately for each laser intensity, since fields of different strength lead to different population of electronic states. However, since this paper focuses more on exploring new methods for simulating HHG than the construction of basis sets, we decided to use the same basis sets for all laser conditions for the sake of simplicity.

The one- and two-electron integrals necessary for the calculations are generated using the Dalton2020.0 software \cite{dalton}. All other calculations, including solving the Hartree-Fock equations, construction and diagonalization of the CI matrices and real-time propagations are performed using a home-made program.

\subsection{Treatment of ionization within the TD-CI approach}

The incompleteness of the used basis sets, both in terms of the radial functions and angular momenta, causes the wavefunction to undergo unphysical reflections when it reaches the area where it cannot be properly described with Gaussian functions. A similar problem is encountered in the grid-based calculations, where the wavefunction is being reflected after reaching the grid boundary. Several methods have been developed to compensate for such an unwanted behavior, including complex absorbing potentials \cite{riss1996,muga2004}, exterior complex scaling \cite{simon1979} and mask functions \cite{krause1992}. All of them rely on eliminating these parts of the wavefuction that cannot be reproduced by the propagation method in use, effectively simulating the ionization process. In our calculations we implement the so-called heuristic lifetime model developed by Klinkusch \textit{et al.} \cite{klinkusch2009a} specifically for the TD-CI propagation scheme. The model relies on the assumption that the CI states above the ionization threshold have finite lifetimes and replaces their energies by complex counterparts,
\begin{equation}
E_k^{\textrm{CI}} \rightarrow E_k^{\textrm{CI}} - \frac{i}{2} \Gamma_k \quad \text{for} \; E_k^{\textrm{CI}} \geq E_0^{\textrm{CI}} + I_p,
\end{equation}
where $\Gamma_k$ is the ionization rate (inverse lifetime) of the $k$-th CI state. The values of $\Gamma_k$ are calculated in the one-electron picture. Assuming that an electron on orbital $a$ with positive energy $\epsilon_a$ has the kinetic energy $\epsilon_a = \frac{1}{2}v^2$, one can introduce an empirical parameter $d_a$ defining the distance the electron can travel before being ionized, and derive the ionization rate $\gamma_a$ of this orbital as its inverse lifetime $\tau_a$,
\begin{equation}
\gamma_a = 1/\tau_a = \theta(\epsilon_a) \sqrt{2\epsilon_a}/d_a,
\end{equation}
where $\theta(x)$ is the Heaviside step function. Therefore the expressions for the ionization rates of CIS and CISD states read
\begin{eqnarray}
\Gamma_k^{\textrm{CIS}} = &&\sum_{i,a} \abs{C^a_{i,k}}^2 \gamma_a, \\
\Gamma_k^{\textrm{CISD}} = &&\sum_{i,a} (\abs{C^a_{i,k}}^2 \gamma_a + \abs{C^{aa}_{ii,k}}^2 2\gamma_a) \\ \nonumber
&&+ \sum_{i,a,b} \abs{C^{ab}_{ii,k}}^2 (\gamma_a + \gamma_b) + \sum_{i,j,a} \abs{C^{aa}_{ij,k}}^2 2\gamma_a \\ \nonumber
&&+ \sum_{i,j,a,b} (\abs{{^\mathrm{A}}C^{ab}_{ij,k}}^2 (\gamma_a + \gamma_b) + \abs{{^\mathrm{B}}C^{ab}_{ij,k}}^2 (\gamma_a + \gamma_b)).
\end{eqnarray}

Klinkusch \textit{et al.} originally considered only one value of $d$ for all orbitals with positive $\epsilon_a$ \cite{klinkusch2009a}. However, Coccia \textit{et al.} \cite{coccia2016a} recently reported that, due to a relatively small number of continuum electronic states obtained with Gaussian basis sets and an uneven spacing between their energy levels, such a choice leads to numerous artifacts in the HHG spectra, like an incorrect description of the harmonic cutoff. Thus, in our version of the model we use two values of $d$. For orbitals with energy smaller than the harmonic cutoff energy predicted by the three-step model, $E_\textrm{cut} = I_p + 3.17 U_p$, we use $d_1$ equal to the semiclassical electron quiver amplitude in an electromagnetic field, $E_0/\omega_0^2$, while for orbitals with energy exceeding $E_\textrm{cut}$ we use $d_2$ = 0.1. The value of $I_p$ is obtained in the Hartree-Fock approximation from the Koopmans' theorem.

It should be noted that our correction to the finite lifetime model is somewhat similar to the correction proposed by Coccia \textit{et al.} \cite{coccia2016a,coccia2019}, who assigned different values of $d$ to whole CI states based on their total energy, rather than to single orbitals. The two approaches are essentially the same for a single-electron system, and give similar ionization rates of CIS states, but the ionization rates of CISD states provided by them may vary significantly. For instance we can consider a state dominated by configurations with double excitations to orbitals with energies just above zero. Since states with a large contribution of doubly excited configurations are usually characterized by high $E_k^{\textrm{CISD}}$, it may occur that the total energy of this state exceeds $E_\textrm{cut}$ and this state is classified by the model of Coccia \textit{et al.} as having an unphysically short lifetime, even though the kinetic energies of both excited electrons are relatively low. Our version of the model is more consistent with the single-electron picture adopted by Klinkusch \textit{et al.}, since it acknowledges that it's the kinetic energy of single \textit{electrons} what limits the harmonic generation and not the total energy of $N$-electron wavefunctions.

\subsection{Grid propagations}

Aside from the basis set propagations, we also solve the time-dependent Schr\"odinger equation for the helium atom on a spatial grid, at the same laser conditions. Since the direct propagation of the six-dimensional wavefunction is currently out of range, we perform the calculations within the single-active-electron (SAE) approximation. In terms of the depiction of the electronic wavefunction this is the crudest method of all applied in this work, as not only it completely neglects all correlation effect, but it also explicitly considers the motion of only one electron, while the other is represented together with the nucleus in the form of a spherically symmetric effective potential. On the other hand, the SAE approximation is known to well reproduce the key features of the HHG spectra \cite{ivanov2009,le2009}. What is even more important, grid projection of the wavefunction provides a very close approximation to the complete orbital basis, which is why the SAE approach is often used as a benchmark for basis set calculations \cite{castro2015,reiff2020,pauletti2021}.

We perform the grid-based propagations using a modified version of QPROP TDSE solver \cite{bauer2006,tulsky2020}. The SAE wavefunction is expanded into partial waves with angular momenta up to $l$=50. The angular dependence of each partial wave is treated analytically using spherical harmonics, while its radial component is discretized on a one-dimensional grid with spacing $\Delta r$=0.1 a.u. The time-dependent Hamiltonian reads
\begin{equation}
\hat H(\vec r,t) = -\frac{\nabla^2}{2} + V_\textrm{SAE}(r) + V_\textrm{CAP} - iA(t)\frac{\partial}{\partial z},
\end{equation}
where the last term is the electron-field interaction operator expressed in velocity gauge through the vector potential A(t) connected to the time-dependent electric field via $E(t)=-\partial_t A(t)$. $V_\textrm{SAE}$ is the SAE potential,
\begin{equation}
V_\textrm{SAE}(r) = -\frac{1+e^{-2.135r}}{r},
\end{equation}
that is known to approximately reproduce the first few ground and excited state energies of the helium atom. $V_\textrm{CAP}$ is the complex absorbing potential (CAP) that simulates the ionization process in a similar manner to the finite lifetime model used in TD-CI calculations. We use CAP in the form developed by Manolopoulos \textit{et al.} \cite{manolopoulos2002}, with absorption parameters $\delta=0.2$ and $k_\mathrm{min}=0.2$. For the consistency with TD-CI propagations the CAP starting position is also placed at the distance from the nucleus equal to the electron quiver amplitude in the laser field.

\subsection{Calculation of the HHG spectra}

After performing real-time propagations we calculate the HHG spectra in the dipole form and in the velocity form, using the Fourier transform, 
\begin{equation}
I_{\mathrm{HHG}}(\omega) = \abs{\frac{1}{t_f-t_i}\int_{t_i}^{t_f}G(t) e^{i\omega t}dt}^2,
\end{equation}
where $t_i$ and $t_f$ are initial and final propagation times, respectively, and $G(t)$ is either the time-resolved dipole moment calculated in the TD-CI picture as
\begin{equation}
\expval{D(t)} = \sum_{k,l} [c_l(t)]^\dagger c_k(t) \langle\Psi_l\vert \hat z \vert\Psi_k\rangle,
\end{equation}
or the dipole velocity, computed by the virtue of the Ehrenfest theorem,
\begin{equation}
\expval{\dot D(t)} = -i \sum_{k,l} [c_l(t)]^\dagger c_k(t) \langle\Psi_l\vert \hat \partial_z \vert\Psi_k\rangle.
\end{equation}

In case of grid propagations, QPROP does only provide the time-resolved dipole moment and not the time-resolved dipole velocity. Therefore we compute the velocity form of the spectra using the approximate relation between the dipole form $I_{\mathrm{HHG,z}}(\omega)$ and the velocity form $I_{\mathrm{HHG,v_z}}(\omega)$:
\begin{equation}
I_{\mathrm{HHG,v_z}}(\omega) \approx \omega^2 I_{\mathrm{HHG,z}}(\omega).
\end{equation}

\section{Results}

Let us start the analysis of the obtained results with some general consideration regarding how the inclusion of double excitations in the CI picture affects the obtained time-independent Hamiltonian eigenspectrum, as it will help us better understand the features of the HHG spectra. We can distinguish three effects stemming from adding doubly excited configurations to the CI wavefunction. The first one is the increase in the total number of states, which is rather self-explanatory given the larger size of the CISD Hamiltonian matrix. As mentioned in the preceding section, CISD states dominated by doubly excited configurations usually have high energies, so one can expect a denser distribution and wider range of states mainly above the ionization threshold. The second effect is the lowering of energies of the CI states that are already approximated by the CIS level of theory (including the ground state) by the values of correlation energy. The third and final effect is the ability of CISD to reproduce states of higher total angular momentum than CIS within the same basis set. This is due to the total angular momentum of a many-electron wavefunction $L$ being a sum of angular momenta of single electrons. In case of the helium atom, CIS incorporates only the configurations in which one electron is excited to an orbital with any $l$ up to $l_\textrm{max}$, while the other remains on the ground state orbital with $l$=0, thus the maximum total angular momentum of the two-electron wavefunction is equal to $l_\textrm{max}$. In CISD on the other hand both electrons can be potentially excited to orbitals with $l$=$l_\textrm{max}$, so the maximum total angular momentum equals $2l_\textrm{max}$. Therefore on the CISD/d-aug-cc-pVQZ+ARO6f level of theory we can obtain states with higher total angular momentum than on the CIS/d-aug-cc-pVQZ+ARO8g level of theory, despite using basis set with less orbital angular momenta. These three effects obviously cannot be strictly separated from each other, but their mutual interplay governs the description of the laser-driven electron dynamics.

\begin{figure}
\centering
\includegraphics[width=.8\linewidth]{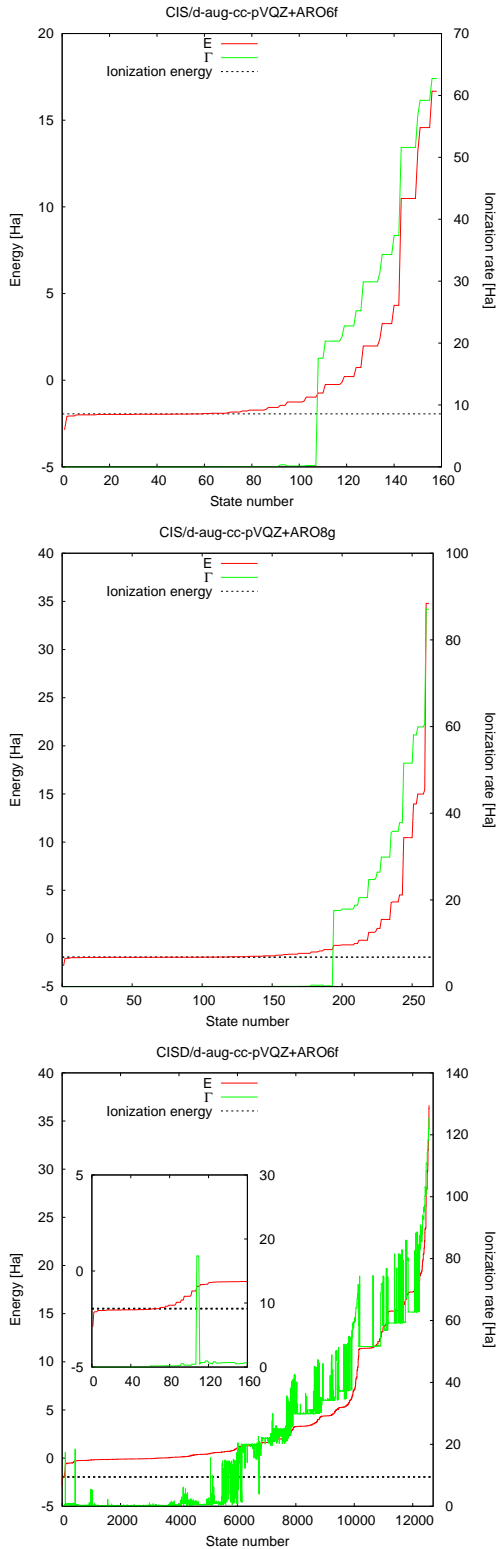}
\caption{Distributions of energies and ionization probabilities of the CI states obtained at the CIS/d-aug-cc-pVQZ+ARO6f, CIS/d-aug-cc-pVQZ+ARO8g and CISD/d-aug-cc-pVQZ+ARO6f level of theory. Inset on the bottom plot shows the distribution for the first 160 CISD states, for an easier comparison with two upper plots}
\label{fig:distros}
\end{figure}

The energies and ionization rates of CI states computed at the CIS/d-aug-cc-pVQZ+ARO6f, CIS/d-aug-cc-pVQZ+ARO8g and CISD/d-aug-cc-pVQZ+ARO6f levels of theory (hereafter CIS/6f, CIS/8g and CISD/6f for short, respectively) are plotted on Fig. \ref{fig:distros}. As predicted, both CIS/6f and CISD/6f calculations provide a similar number of about 60 bound states, while the number of high-energy continuum states in the former method is smaller by two orders of magnitude than in the latter. Contrariwise, the addition of more ARO functions to the basis set results in much higher density of states close to the ionization threshold. The ionization rates of CIS states calculated in both basis sets rise consecutively with the state number. There is a sharp increase in $\gamma$ about 105th and 190th state in CIS/6f and CIS/8g eigenspectrum, respectively, indicating that the states beyond these points contain excitations to orbitals with $\epsilon > E_\textrm{cut}$. The same dependency for the CISD states is far more oscillatory, which illustrates the effect discussed in the previous section: states dominated by configurations in which one electron is excited to a high-energy continuum orbital and the other is excited to a bound orbital or not excited at all (e.g. states with high total ionization rates), and states dominated by configurations with double excitations to low-energy continuum orbitals (e.g. states with low total ionization rates) may have similar total energies. 

\begin{figure*}
\centering
\includegraphics[width=1.\linewidth]{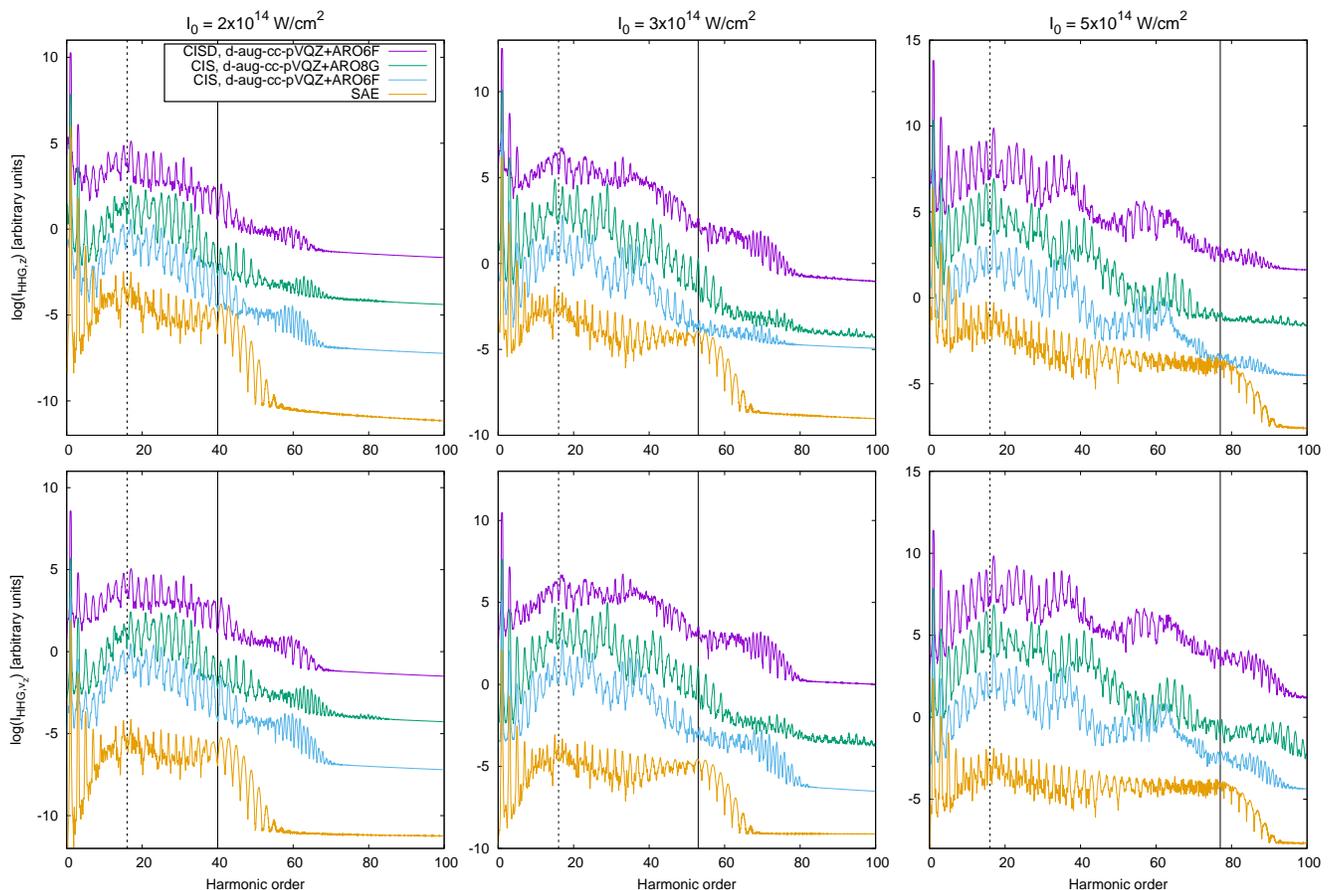}
\caption{HHG spectra of the helium atom at three examined laser intensities, in the dipole form (top row) and in the velocity form (bottom row). The TD-CIS/6f, TD-CIS/8g and TD-CISD/6f spectra are upshifted along the $y$-axis by the values of 2, 4 and 6, respectively, for sake of clarity. The theoretical cutoff position predicted by the three-step model (solid line) and the harmonic corresponding to the ionization energy (dashed line) are denoted on each plot}
\label{fig:spectra}
\end{figure*}

The HHG spectra obtained from the real-time propagations are presented on Fig. \ref{fig:spectra}. We start the comparison with some remarks on the grid-based results. The SAE approximation is shown to well reproduce the overall shape of the spectra in both forms and at all laser conditions. The intensities of the consecutive peaks in the harmonic plateaux decrease slowly and almost monotonically with the harmonic number. Each plateau ends with a sharp cutoff, after which the HHG intensity abruptly drops by a several orders of magnitude. The cutoff position is in a good agreement with the value predicted by the three-step model at every examined laser intensity. On the other hand, the harmonic peaks are less distinct and more contaminated by numerical noises than the ones on the spectra obtained from TD-CI propagations. The intensity of these noises rises with the harmonic number, so that at the spectrum corresponding to intensity $5 \times 10^{14}$ W/cm\textsuperscript{2} the peaks before the cutoff cannot be distinguished, suggesting a rather poor representation of highly excited and continuum states in the SAE picture. It should be noted that these artifacts could not be eliminated by adding more partial waves to the wavefunction expansion or by decreasing the grid spacing, implying the limitations of the SAE model itself.

At the laser intensity $2 \times 10^{14}$ W/cm\textsuperscript{2} all three TD-CI methods perform fairly good at reproducing the harmonic peaks in the plateau, as well as at predicting the cutoff positon. However, on deeper analysis, TD-CISD/6f produces the spectrum of arguably the best quality. It provides the smallest amount of excess harmonic peaks behind the cutoff and the least numerical noises in the harmonic background. Moreover, the plateaux on both TD-CIS-based spectra are characterized by a triangular envelope, with the intensities of the peaks increasing up to the point corresponding to the ionization threshold, and then decreasing until reaching the cutoff. The TD-CISD-based spectrum provides a more constant intensity of the plateau, resembling that in the SAE-based spectrum. This effect is more noticeable in the velocity form.

At the laser intensity $3 \times 10^{14}$ W/cm\textsuperscript{2} TD-CIS/6f is still able to reproduce most of the harmonic plateau, although its performance  starts to decline, with the cutoff harmonic peak being just distinguishable in the dipole form of the TD-CIS/6f spectrum, and the cutoff position being placed beyond the value predicted by the three-step model in the velocity form. A much better description of the cutoff region is achieved with TD-CIS/8g, although at the cost of some numerical artifacts in the background, visible particularly in the velocity form of the spectrum.

Surprisingly, TD-CISD/6f provides a poorer description of the HHG plateau at this intensity than both TD-CIS/6f and TD-CIS/8g, with the peaks being less distinct especially in the region between about the 40th harmonic and the cutoff. A potential explanation for this observation is following. On the basis of the differences between the TD-CIS/6f and TD-CIS/8g spectrum we can infer that the states with $L > 3$ are necessary to better reproduce the highest order peaks in the plateau. In CISD/6f such states can be constructed, but only from configurations with double excitations to orbitals with $l \leq 3$.
However, the ARO functions are optimized in the one-electron picture, so it is natural that an ARO shell with angular momentum $l$ is best suited for describing states with only one electron excited to orbital with the same angular momentum.
Therefore, it is inevitable that the properties of states with $L > 3$, composed solely of doubly excited configurations, are approximated less accurately than the properties of states with $L \leq 3$, comprised of both singly and doubly excited configurations.
This applies in particular to the dipole moment expectation value, from which the HHG spectrum is derived.
This imbalanced description of different angular momenta within the time-dependent wavefunction is a source of additional artifacts, which do not occur if the states with $L > 3$ are not included at all.
On the other hand, it is worth noticing that of all TD-CI-based spectra in the dipole form a decrease of background intensity beyond the cutoff is present only in the TD-CISD/6f one. Given that such a decrease is observed in both forms of all grid-based spectra, this suggests a possible key role of states with high angular momenta in the depiction of this feature.

At the laser intensity $5 \times 10^{14}$ W/cm\textsuperscript{2} the performance of TD-CIS/6f breaks down, as the HHG intensity abruptly falls off after the 63th harmonic and no higher order peaks are present in the spectrum. This indicates the lack of CIS states with energy or angular momentum high enough to describe the laser-driven electron dynamics in such an intense field.
TD-CIS/8g on the other hand is still able to reproduce all harmonic peaks in the plateau at this intensity, with the performance comparable to that of TD-CIS/6f at intensity $3 \times 10^{14}$ W/cm\textsuperscript{2}. 
TD-CISD/6f provides a rather similar picture to TD-CIS/6f, with only minor differences. In the first half of the plateau, encompassing harmonics from first to 40th, we observe some improvement in terms of more constant intensity of consecutive peaks than in both TD-CIS-based spectra. This effect is similar to the one observed at intensity $2 \times 10^{14}$ W/cm\textsuperscript{2}, and is a strong hint for the role of electronic correlation in HHG. This is because these lower order peaks are generated mostly by transitions to states with relatively low excitation energies, which are approximately described by all three levels of theory, but in TD-CISD/6f this description is corrected by including doubly excited configurations. In the second half of the plateau (region between the 40th harmonic and the cutoff) the peaks that are visible clearly in the TD-CIS/6f spectrum are less distinct here, due to the same reason as in case of the intensity $3 \times 10^{14}$ W/cm\textsuperscript{2}. However, it can be noticed that the region between the 63th harmonic and the cutoff the HHG intensity is slightly uplifted compared to the TD-CIS/6f spectrum, indicating transitions to states with $L > 3$, that are impossible to realize on the TD-CIS/6f level of theory. Nevertheless, the poor description of these states using only doubly excited configurations makes it unfeasible to distinguish individual harmonic peaks.

\section{Conclusion}

In this work we have compared the performance of three real-time time-dependent wavefunction propagation methods: TD-CI with single excitations, TD-CI with single and double excitations, and the single-active-electron approximation, in the modeling of high harmonic generation in the helium atom exposed to laser fields of extreme intensity. The first two approached rely on expanding the time-dependent wavefunction in the predefined basis set comprised of Gaussian functions, while the third is grid-based. The full HHG spectra obtained from TD-CISD calculations are to our knowledge the first to be reported in the literature.

First of all, the results presented in this paper may serve as an additional proof of applicability of the ARO Gaussian functions in the modeling of HHG, extending our previous study \cite{wozniak2021}. We have shown on the example of the helium atom that Dunning basis sets supplemented with ARO functions allow for studying the laser-driven electron dynamics in many-electron systems in the CI framework. Also, it has been proven that subsequent addition of ARO shells with higher angular momenta, as well as augmenting the number of functions per angular momentum makes it possible to simulate the optical response in laser fields of increasing intensity.

This work was mainly motivated by the question of whether the explicit inclusion of electron correlation via the incorporation of doubly excited configurations in the time-dependent CI wavefunction has a significant contribution to the description of HHG. After a careful analysis of the obtained HHG spectra we are able to reach the conclusion that the addition of double excitations can indeed noticeably improve the TD-CI picture of HHG, but only when the basis set in which the calculations are performed already provides qualitatively acceptable results at the TD-CIS level of theory. This is best seen at the intensity $2 \times 10^{14}$ W/cm\textsuperscript{2}. The eigenspectrum of CIS/6f already contains all the singly excited states necessary for an approximate description of the laser-driven wavefunction, and the role of double excitations in CISD/6f is merely to complement this description by including the correlation between the electron that becomes ionized and the one that remains in the residual ion. The effects of this complementation are perceptible in both an improved overall shape of the HHG plateau and a better description of the cutoff region, and are an indisputable evidence for the contribution of electronic correlation in the HHG process. On the other hand, when the basis set lacks the functions of angular momenta high enough to describe the electron dynamics, augmenting the configurational basis by including double excitations cannot compensate this shortcoming and the correct picture can only be regained by expanding the orbital basis, as seen at higher laser intensities. This is because the states exclusive to the CISD eigenspectrum are better suited to describe double excitations and double ionization processes, while in HHG the contribution of single ionization obviously prevails over the contribution of double ionization. Thus, performing TD-CISD calculations in an inadequate basis set brings little to no improvement over the TD-CIS picture (as in the case of intensity $5 \times 10^{14}$ W/cm\textsuperscript{2}) or can even worsen it (as in the case of intensity $3 \times 10^{14}$ W/cm\textsuperscript{2}). Moreover, although CISD provides a far higher overall number of CI states, the majority of them lay far above the ionization threshold, so their presence is not expected to contribute much to the HHG process. On the contrary, expanding the number of ARO functions in the basis set leads to significant increase in the number of states just near the ionization threshold, which are believed to play a major role in HHG.

Unfortunately, such a conclusion limits the applicability of the TD-CISD approach to systems for which non-correlated calculations in a comparatively small basis set already provide a qualitatively correct results, i.e. atoms or molecules with few electrons, subjected to fields of relatively low intensity. Otherwise, incrementing the number of electrons or the laser intensity means that larger basis sets have to be used in order to capture the electronic correlation effects, which in turn makes the TD-CISD calculations impractical due to their high computational complexity. There are however some potential ways to bypass this restriction. For example, one can truncate the CISD excitation space by removing the highest lying eigenstates, which have negligible contribution to the electron dynamics. Another solution may be to perform calculations for larger atoms or molecules using the effective core potential method, which allows for limiting the number of explicitly treated electrons to valence shell only. Both approaches significantly reduce the order of time propagation matrix equations and will be explored in future works.

This paper also introduces and test a new correction to the finite lifetime ionizaton model of Klinkusch \textit{et al.} \cite{klinkusch2009a}. As shown by our results, it ensures a fairly good agreement of the cutoff position with the value predicted by the three-step model even at extremely high laser intensities. 
\section*{Acknowledgments}

We acknowledge support from
the National Science Centre, Poland (Symfonia Grant No. 2016/20/W/ST4/00314).
M. L. also acknowledges support from
ERC AdG NOQIA;
Agencia Estatal de Investigación (R\&D project CEX2019-000910-S, funded by MCIN/AEI/10.13039/501100011033, Plan National FIDEUA PID2019-106901GB-I00, FPI, QUANTERA MAQS PCI2019-111828-2, Proyectos de I+D+I “Retos Colaboración” RTC2019-007196-7);
Fundació Cellex;
Fundació Mir-Puig;
Generalitat de Catalunya through the CERCA program, AGAUR Grant No. 2017 SGR 134, QuantumCAT\textbackslash U16-011424, co-funded by ERDF Operational Program of Catalonia 2014-2020;
EU Horizon 2020 FET-OPEN OPTOLogic (Grant No 899794);
Marie Sk{\l}odowska-Curie grant STREDCH No 101029393;
“La Caixa” Junior Leaders fellowships (ID100010434) and EU Horizon 2020 under Marie Sk{\l}odowska-Curie grant agreement No 847648 (LCF/BQ/PI19/11690013, LCF/BQ/PI20/11760031, LCF/BQ/PR20/11770012, LCF/BQ/PR21/11840013).

\bibliography{apssamp}

\end{document}